\documentclass[aps,preprint,nofootinbib,prc]{revtex4}

\usepackage{graphicx}
\usepackage{dcolumn}
\usepackage{bm}

\begin{document}

\title{Pre-neutron emission mass distributions for low-energy neutron-induced actinide fission}

\author{Xiaojun Sun}
\email{sxj0212@gxnu.edu.cn}
\author{Chenggang Yu}
\author{Ning Wang}
\email{wangning@gxnu.edu.cn}
\affiliation{Department of Physics, Guangxi Normal University, Guilin 541004, P. R. China}

\date{\today}

\begin{abstract}

According to the driving potential of a fissile system, we propose
a phenomenological fission potential for a description of the
pre-neutron emission mass distributions of neutron-induced
actinide fission. Based on the nucleus-nucleus potential with the
Skyrme energy-density functional, the driving potential of the
fissile system is studied considering the deformations of nuclei.
The energy dependence of the potential parameters is investigated
based on the experimental data for the heights of the peak and
valley of the mass distributions. The pre-neutron emission mass
distributions for reactions $^{238}$U(n, f), $^{237}$Np(n, f),
$^{235}$U(n, f), $^{232}$Th(n, f) and $^{239}$Pu(n, f) can be
reasonably well reproduced. Some predictions for these reactions
at unmeasured incident energies are also presented.
\end{abstract}

\pacs{24.75.+i, 25.85.Ec, 25.40.-h} \keywords{neutron-induced
actinide fission, pre-neutron emission mass distribution, fission
potential } \maketitle

\section{introduction}

Nuclear fission is a field of very intense studies in the recent
decade
\cite{moller2001,moller2004,madland2006,Lammer2008,moller2009,Schmidt2010,Randrup2011}.
One of the most interesting characteristics of neutron-induced
fission is the huge difference of the mass distribution of the
fission fragments for different nuclei and the dramatic change of
the mass distribution with the variation of the incident energies
of neutron. It is still far from clear how the parent nucleus
transforms into a variety of daughter pairs. The highly excited
primary fission fragments, whose mass distributions are called
pre-neutron emission mass distributions, are de-excited by the
emission of prompt neutrons, and followed by prompt $\gamma$-rays
to form the primary fission products. The primary fission products
are usually highly neutron-rich and unstable, and gradually evolve
to the secondary fission products through the emission of delayed
neutrons and the radioactive $\beta$-decay. We will focus on the
study of the pre-neutron emission mass distribution of the primary
fission fragments in this work. The precise calculation of the
pre-neutron emission mass distributions is of great importance for
understanding the fission process and for describing the yields of
the fission products.

The measured mass distribution of the fission fragments can be
reasonably well reproduced with some empirical approaches or some
systematical methods. Liu \textit{et al}. had developed
systematics of mass distributions for neutron-induced $^{238}$U
fission \cite{liutingjin200801} and of independent yields for
neutron-induced $^{235}$U fission \cite{snc2007}. They also
presented the evaluation data \cite{liutingjin200802} and the
adjusted data \cite{liutingjin200803} for several actinides.
Katakura \cite{katakura2008, katakura2003} and Wahl
\cite{wahl2008} had fitted the experimental data for actinide
nuclei with 3 to 7 Gaussian functions. Kibkalo's phenomenological
model was designed to study the dependence of the mass
distribution on the transferred angular momentum, and was later
adapted for predictions of fission yields \cite{kibkalo2008}. A
new systematics for fragment mass yields of target nuclei from Th
to Bk at incident particle energies between 5 and 200 MeV was
developed by Gorodisskiy \textit{et al}. through independent
fission modes \cite{Gorodisskiy2008}. The mass distributions
predicted with the systematical methods mentioned above are
generally described by a series of Gaussians, and the model
parameters are obtained through fitting the experimental data.
However, available experimental data for energy-dependent
neutron-induced fission yields, especially the pre-neutron
emission mass distributions, are not sufficient enough for the
development of global systematics, which results in great
difficulties for the predictions of the mass distributions at
unmeasured energies.

For a microscopic description of the mass distribution of nuclear
fission, the precise calculation of the potential energy surface
seems to be required. Unfortunately, the microscopic calculation
of the potential energy surface of a fissile system is very
complicated and time-consuming. Some phenomenological approaches
are still required for the quantitative description of the energy
dependence of the mass distribution at present. It is known that
the shell and pairing effects play a key role for the fission and
quasi-fission process. It is found that the quasi-fission mass
distribution in fusion reactions leading to the synthesis of
super-heavy nuclei can be reasonably well described by the driving
potential in the di-nuclear system (DNS) model
\cite{Adamian2003,Kalandarov2011}, since the shell effects of
reaction system are effectively involved in the driving potential
via the Q-value of the system. It is therefore interesting to
investigate the mass distribution of a fissile system based on its
driving potential. Because the deformation effect of a nuclear
system influences the process of fusion and fission
\cite{Gupta2005}, it is expected that the deformations of nuclei
play a role for a reliable calculation of the corresponding
driving potential.

In this work, we attempt to propose a simplified fission potential
with a few well-determined parameters for quantitatively
describing the pre-neutron emission mass distributions of
neutron-induced actinide fission, by combining the corresponding
driving potential of the fissile system. This paper is organized
as follows: In Sec. II, we briefly introduce the calculation of
the driving potential of a fissile system. In Sec. III, we
introduce the fission potential and its parameters. In Sec. IV,
the comparisons between the predicted results and the measured
data of the pre-neutron emission mass distributions for the
reactions $^{238}$U(n, f),
 $^{237}$Np(n, f), $^{235}$U(n, f), $^{232}$Th(n, f) and  $^{239}$Pu(n, f)  are presented.
Finally, the summary and discussion is given in Sec.V.

\section{driving potential of a fissile system}

Assuming that a compound nucleus separates into a pair of nuclei
in the fission process,
\begin{equation}
(A_{\rm CN},Z_{\rm CN})\rightarrow (A_1,Z_1) + (A_2,Z_2),
\end{equation}
the corresponding Q-value of the system can be expressed as,
\begin{equation}
Q=E(A_{\rm CN},Z_{\rm CN})-E(A_1,Z_1)-E(A_2,Z_2).
\end{equation}
Here, $E (A_i,Z_i)$ denotes the energy of a nuclear system with mass $A_i$
and charge $Z_i$. For a description of the potential energy
surface of a fissile system around scission point, the dinuclear
system (DNS) concept may be used, omitting the excitation energies
of the fission fragments for simplicity. According to the DNS
concept, each fission fragment at the scission point retains its
individuality in the evolution of the DNS. This is a consequence
of the influence of the shell structure of the partner fragments
since the collective kinetic energy of the fission fragments is
low around the scission point. Based on the DNS concept, the
driving potential of a fissile system is expressed as
\begin{equation}
D =Q + B_0
\end{equation}
in this work. $B_0$ denotes the Coulomb barrier height in the
interaction potential between the fragment pair. In this work, the
interaction potential $V(R)$ at a center-to-center distance $R$ is
calculated by using the Skyrme energy-density functional together
with the extended Thomas-Fermi (ETF) approximation \cite{Liu06}.

\begin{figure}[!ht]
\begin{center}
\includegraphics[angle=0,width=0.85\textwidth]{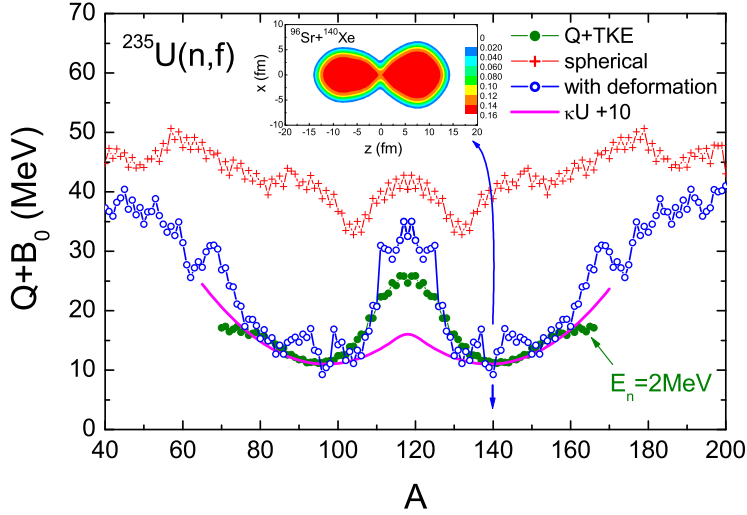}
\caption{(Color Online)  Driving potential for $^{235}$U(n, f). The
solid circles denote the sum of the Q-value and the average total
kinetic energy $\overline {\rm TKE}$ of fission fragments. The
crosses and open circles denote the driving potential $D$ without
and with the deformations of nuclei being taken into account,
respectively. The solid curve denotes the results of $\kappa U +10$
with $\kappa=1$ MeV. $U(A)$ denotes the empirical fission potential
which will be discussed in next section. The sub-figure shows the
corresponding density contour plots of a typical primary fragment
pair $^{96}$Sr+$^{140}$Xe.}
\end{center}
\end{figure}

Fig. 1 shows the driving potential for  $^{235}$U(n, f). The solid
circles denote the sum of the Q-value and the average total
kinetic energy $\overline {\rm TKE}$ of fission fragments at an
incident energy of 2 MeV. One sees that there exists a valley at
$\sim 140$ for the mass number of the heavy fission fragments. The
open circles denote calculated driving potential with the
deformations of nuclei being taken into account. Here, the
deformations of nuclei at their ground state are taken from the
calculations of the finite range droplet model \cite{Moll95}. We
consider the tip-tip orientation and simultaneously consider the
dynamical octupole deformation of fragments (empirically set
$|\beta_3|=0.08$) in the fission process around the scission
point, since one obtains a lower Coulomb barrier at the tip-tip
orientation. We note that the measured data for $Q+ \overline {\rm
TKE}$ are roughly reproduced by the calculated driving potential.
The crosses denote the results without the deformations of nuclei
being taken into account. From the comparison, one learns that the
deformations of fragments play an important role for a reasonable
description of the fission mass distribution and the total kinetic
energy of fragments. The solid curve denotes the result from an
empirical fission potential which will be discussed in next
section.

\section{fission potential and its parameters}

For a more quantitative description of the mass distribution of
primary fission fragments, we further propose an empirical fission
potential considering the calculated driving potential. One
expects that the pre-neutron emission mass distribution, i.e., the
mass dependence of the primary fission fragments are strongly
dependent on a corresponding fission potential. We assume that the
pre-neutron emission mass distributions of low-energy
neutron-induced actinide fission can be approximated described by
using a simplified fission potential $U(A)$,
\begin{equation}
P(A)=C \exp[-U(A)].
\end{equation}
Where $C$ is the normalization constant, and the variable $A$
denotes the mass number of the primary fragment. Considering the
double-humped mass distributions of low-energy neutron-induced
actinide fission, we describe the phenomenological fission
potential $U(A)$ by using three harmonic-oscillator functions,
i.e.,
\begin{equation}\label{eq2}
    U(A)= \left\{\begin{array}{l l}
        \displaystyle u_1(A-A_1)^2               & A\leq a      \\
        \displaystyle -u_0(A-A_0)^2+R            & a\leq A\leq b \\
        \displaystyle u_2(A-A_2)^2               & A\geq b.      \\
    \end{array}\right.
\end{equation}
Where, $A_1$, and $A_2$ are the positions for the peaks of the
light and heavy fragments of the pre-neutron emission mass
distributions, respectively. According to the calculated driving
potentials, we find that the valley for the mass number of heavy
fragment locates $A_2\approx 140$ for neutron-induced actinide
fission in general. Therefore, we set $A_2=140$ in the calculation
for simplicity. We have checked that the calculated mass
distribution of fission fragments do not change appreciately if
the value of $A_2$ is slightly changed. $A_0=A_{\rm CN}/2$ denotes
the corresponding position for symmetric fission. Here, $A_{\rm
CN}$ is the mass number of the fissile nucleus. Considering that
the fission potential is a smooth function, the coefficients in
Eq.(5) can be derived as
\begin{eqnarray}\label{eq3}
u_0 &=& \frac{R}{(A_0-a)(A_0-A_1)},\nonumber\\
u_1 &=& \frac{R}{(A_0-A_1)(a-A_1)},\nonumber\\
u_2 &=& \frac{R}{(A_2-A_0)(A_2-b)},
\end{eqnarray}
with $A_1=A_{\rm CN}-A_2$ and $b = \displaystyle
\frac{(A_0-a)(A_0-A_1)}{A_2-A_0}+A_0$. The potential parameters $a$
and $R$ will be discussed later.

The total mass distributions of the binary fission fragments
should be normalized to $200\%$. The normalization constant $C$
can therefore be analytically expressed as
\begin{equation}\label{eq4}
C
=\frac{200\%}{\int_0^{\infty}\exp[{-U(A)}]dA}=\frac{200\%}{I_0+I_1+I_2},
\end{equation}
with
\begin{eqnarray}\label{eq5}
I_0 &=& \frac{\sqrt{\pi}e^{-R}}{2\sqrt{u_0}} \{   {\rm erfi}[(A_0-a)\sqrt{u_0} \;]+ {\rm erfi}[(b-A_0)\sqrt{u_0} \;]   \} , \nonumber\\
I_1 &=& \frac{\sqrt{\pi}}{2\sqrt{u_1}}\{\textmd{erf}[(a-A_1)\sqrt{u_1} \;]+\textmd{erf}[A_1\sqrt{u_1} \;]\}, \nonumber\\
I_2 &=&
\frac{\sqrt{\pi}}{2\sqrt{u_2}}\{1+\textmd{erf}[(A_2-b)\sqrt{u_2}
\;]\}.
\end{eqnarray}
Where ${\rm erf}(x)$ and ${\rm erfi}(x)$ denotes the error
function and imaginary error function, respectively. We also
assume that $P(A_1)=P(A_2)$, i.e., for the pre-neutron emission
mass distributions the height of the peak of
 the light fragments equals to that of the heavy fragments. The
 parameter $a$ can be uniquely determined by the normalization
 constant $C$ and $P(A_1)$.

\begin{figure}[!ht]
\begin{center}
\includegraphics[angle=0,width=0.7\textwidth]{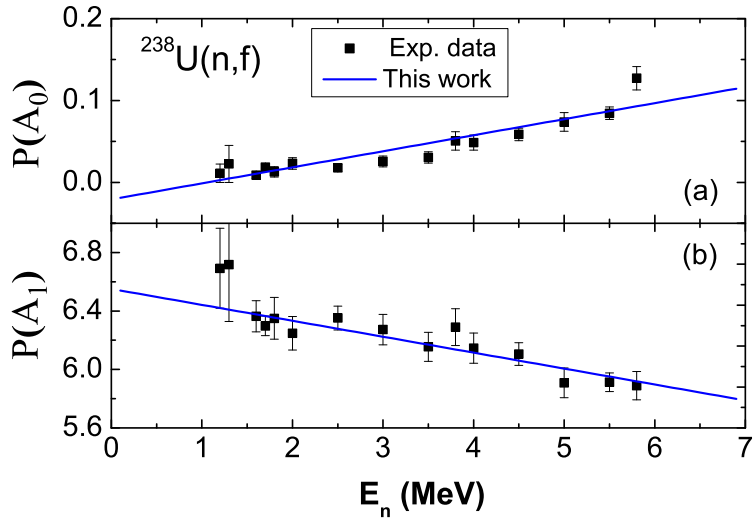}
\caption{\label{VT}(Color Online)  Values of peak $P(A_1)$ and
valley $P(A_0)$ of the pre-neutron emission mass distributions for
reaction $^{238}$U(n, f)  as a function of incident energy of
neutron. The data are taken from Ref. \cite{vive2000}. The solid
lines denote the results in this work.}
\end{center}
\end{figure}

The potential parameter $R$ is defined as
\begin{eqnarray}\label{eq7}
R=\ln\frac{P(A_1)}{P(A_0)}.
\end{eqnarray}
We find that the heights of the peak and valley of the mass
distributions change linearly with the low incident energies of
neutron in general. The energy dependence of $P(A_0)$  and
$P(A_1)$ are written as,
\begin{eqnarray}\label{eq8}
P(A_0)&=&\alpha_0+\beta_0 E_n, \nonumber\\
P(A_1)&=&\alpha_1+\beta_1 E_n.
\end{eqnarray}
Here, $E_n$ denotes the incident energy of neutron. The parameters
$\alpha_0$, $\beta_0$, $\alpha_1$ and $\beta_1$ are finally
determined by the experimental data for $P(A_0)$  and $P(A_1)$.
The potential parameters for different reaction systems are listed
in Table I.

\begin{table}
\caption{The potential parameters adopted in this work.}
\begin{tabular}{lccccc}
\hline\hline
  reaction   ~~~~    &  ~~~~$\alpha_1$~~~~  &~~~~$\beta_1$ ~~~~  &   ~~~~$\alpha_0$  ~~~~      & ~~~~$\beta_0$  ~~~~      \\ \hline
  $^{232}$Th(n, f) &   7.0900         &-0.0631         &     -0.0656            &0.0350             \\
  $^{235}$U(n, f)  &   6.6550         &-0.0952         &     -0.0042            &0.0237             \\
  $^{237}$Np(n, f) &   6.4132         &-0.1008         &     -0.0113            &0.0338             \\
  $^{238}$U(n, f)  &   6.5508         &-0.1090         &     -0.0208            &0.0196             \\
  $^{239}$Pu(n, f) &   6.1293         &-0.1517         &      0.0067            &0.0343             \\
 \hline\hline
\end{tabular}
\end{table}

\section{results}


\begin{figure}
\begin{center}
\includegraphics[angle=0,width=0.7\textwidth]{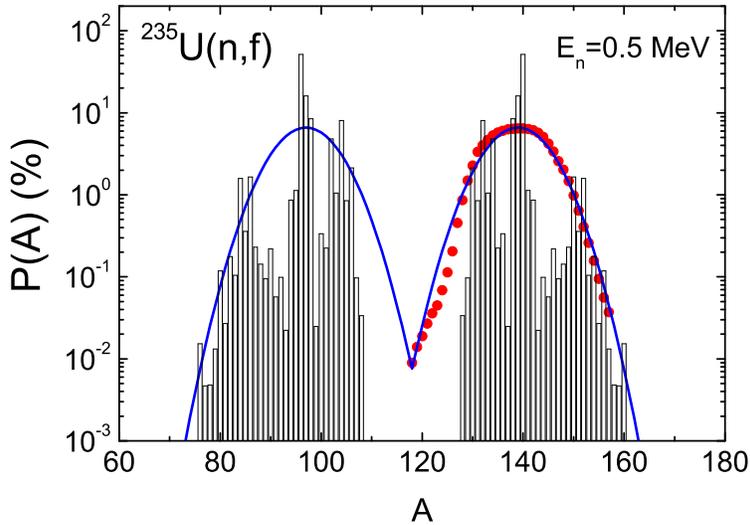}
\caption{\label{VT}(Color Online)  Pre-neutron emission mass
distributions at an incident energy $E_n=0.5$ MeV for reaction
$^{235}$U(n, f). The scattered symbols denote the experimental
data which are taken from Ref. \cite{Djachenko1969}.  The solid
curve and bars denote the calculated results with the empirical
fission potential $U(A)$ and the driving potential, respectively.
}
\end{center}
\end{figure}

\begin{figure*}[!ht]
\begin{center}
\includegraphics[angle=0,width=0.85\textwidth]{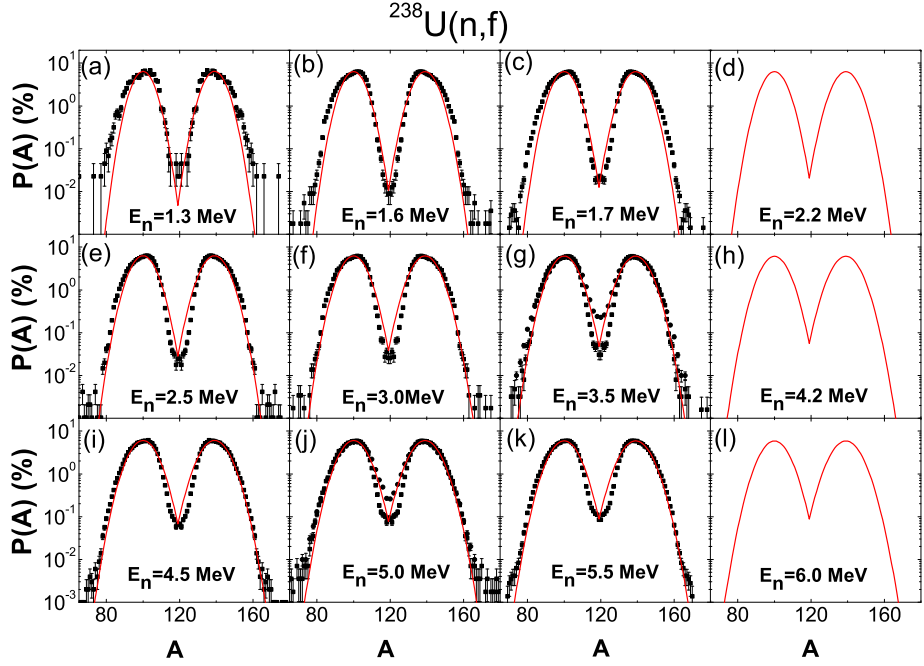}
\caption{\label{u238}(Color Online) Pre-neutron emission mass
distributions at incident energies from 1.3 MeV to 6 MeV for
reaction $^{238}$U(n, f). The scattered symbols denote the
experimental data which are taken from Ref. \cite{vive2000}
(squares) and Ref. \cite{zoller1995} (circles), respectively. The
solid curves denote the calculated results in this work. (d), (h)
and (l) show the predicted results at three unmeasured energies
2.2, 4.2 and 6.0 MeV, respectively.}
\end{center}
\end{figure*}

\begin{figure*}[!ht]
\begin{center}
\includegraphics[angle=0,width=0.85\textwidth]{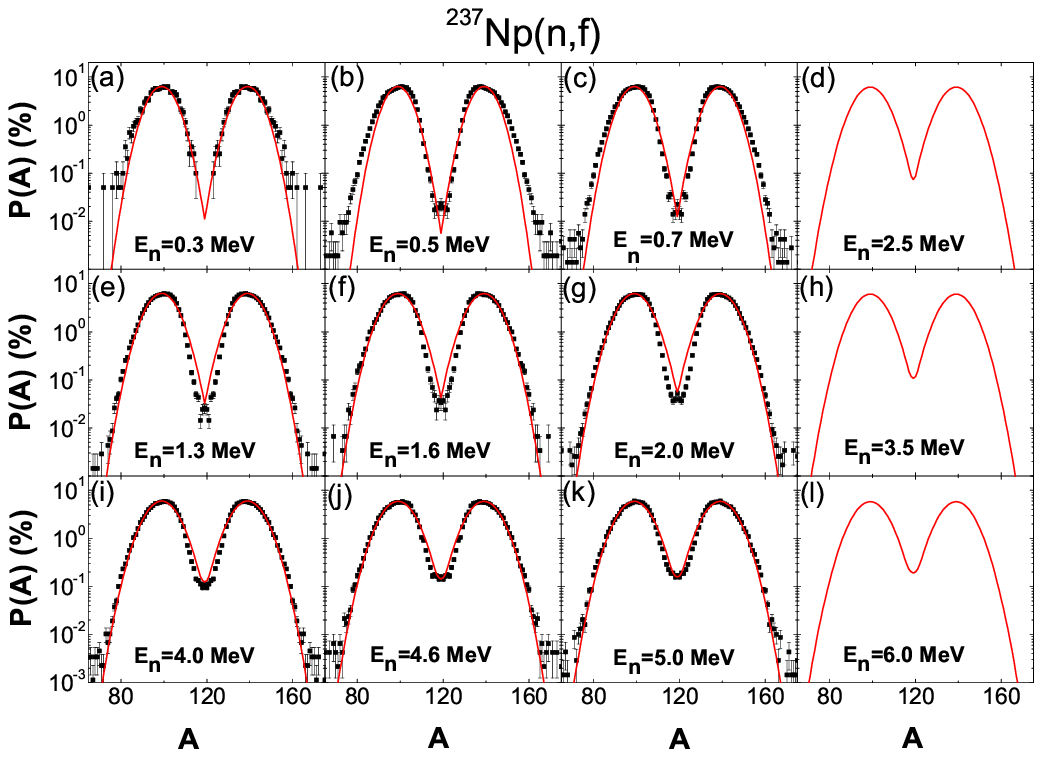}
\caption{\label{Np237}(Color Online) The same as Fig. \ref{u238},
but for reaction $^{237}$Np(n, f)  at incident energies from 0.3
 to 6.0 MeV. The experimental data (square dot) are taken from
Ref. \cite{Hambsch2000}.}
\end{center}
\end{figure*}

\begin{figure*}[!ht]
\begin{center}
\includegraphics[angle=0,width=0.9\textwidth]{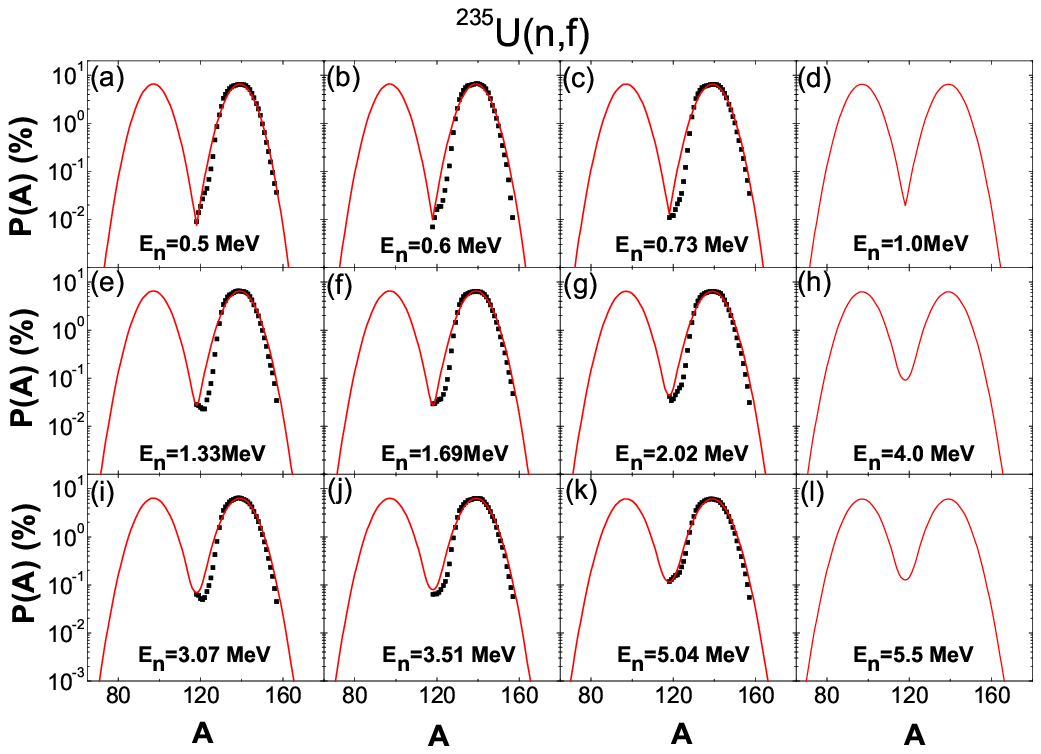}
\caption{\label{u235}(Color Online) The same as Fig. \ref{u238},
but for reaction $^{235}$U(n, f)  at incident energies from 0.5
  to 5.5 MeV. The experimental data (square dot) are taken from
Ref. \cite{Djachenko1969}.}
\end{center}
\end{figure*}

\begin{figure*}[!ht]
\begin{center}
\includegraphics[angle=0,width=0.9\textwidth]{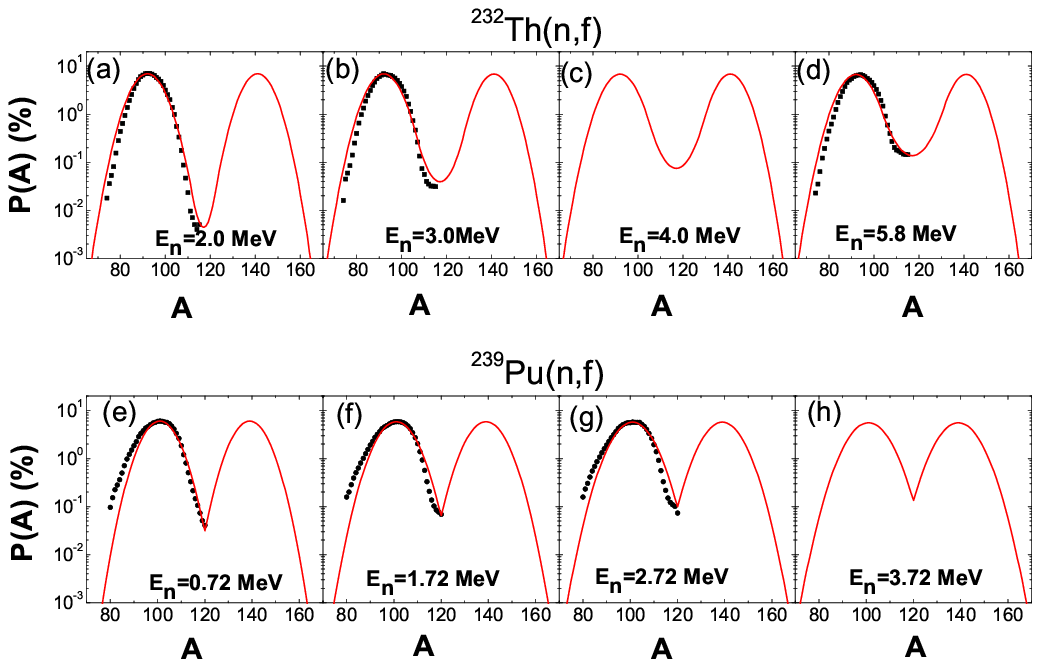}
\caption{\label{ThPu}(Color Online) The same as Fig. \ref{u238},
but (the upper panel) for reaction $^{232}$Th(n, f)  and (the
lower panel) for reaction $^{239}$Pu(n, f) at different incident
energies. The experimental data are taken from Refs.
\cite{sergachev1968, akimov1971}.}
\end{center}
\end{figure*}

In this work, we first investigate the energy dependence of the
potential parameters. In Fig. 2, we show the values of $P(A_1)$
and $P(A_0)$ in the pre-neutron emission mass distributions for
the reaction $^{238}$U(n, f) as a function of the incident energy
of neutron. One can see that the values of $P(A_1)$ and  $P(A_0)$
linearly change with the incident energy in general, which could
provide us with some useful information for calculating the
pre-neutron emission mass distributions at unmeasured energies of
$E_n=1 \sim 6$ MeV. From the potential parameters listed in Table
I, one can see that the values of $\alpha_1$ and $\beta_1$
decrease with the mass number of the fissile nuclei in general. We also note that the
potential parameters, such as $A_2$, $P(A_1)$ and $P(A_0)$,  are
different in different models and the energy dependence of $A_2$ is weak.

In Fig. 3, we show the comparison of the calculated pre-neutron
emission mass distributions $P(A)$ for $^{235}$U(n, f) with the
the empirical fission potential and the driving potential,
respectively. The solid curve denotes the result with $U(A)$ in
Eq.(4). The bars denote the corresponding result with the obtained
driving potential in Fig. 1. Here, the mass distribution is
roughly estimated by using a formula $P(A)\propto \exp(-D/\kappa)$ based
on the obtained driving potential $D$ with $\kappa=1$ MeV and considering the normalization.
One sees the the positions and
widths of the peaks for mass distribution can be reasonably well
reproduced with the driving potential. The large fluctuation is
due to that the temperature dependence of nuclear structure effect
is not considered yet. With the empirical fission potential, the
description of the measured mass distribution for the primary
fission fragments in $^{235}$U(n, f) can be significantly
improved.

In Fig. 4, we show the pre-neutron emission mass distributions
$P(A)$ at incident energies from about 1 to 6 MeV for the reaction
$^{238}$U(n, f). The scattered symbols denote the experimental
data which are taken from Ref. \cite{vive2000} (squares) and from
Ref. \cite{zoller1995} (circles), respectively. The solid curves
denote the calculated results in this work. The potential
parameters $\alpha_0$, $\beta_0$, $\alpha_1$ and $\beta_1$
adopted in the calculations are listed in Table I. Fig. 5, Fig. 6
and Fig. 7 show the pre-neutron emission mass distributions $P(A)$
for the reactions $^{237}$Np(n, f), $^{232}$Th(n, f) and
$^{239}$Pu(n, f), respectively. One can see that the experimental
data can be reproduced reasonably well, which indicates that the
fission potential proposed in this work is reasonable. For some
unmeasured energies, we also present the predictions from this
approach.

\section{summary and discussion}

In this work, we proposed a phenomenological fission potential
based on the corresponding driving potential for quantitatively
describing the pre-neutron emission mass distributions of
neutron-induced actinide fission at incident energies of neutron
at a few MeV. Based on the nucleus-nucleus potential with the
Skyrme energy-density functional, the driving potential of the
fissile system is studied considering the deformations of nuclei.
The measured data for the sum of the Q-value and the average total
kinetic energy $\overline {\rm TKE}$ of fission fragments at an
incident energy of 2 MeV in $^{235}$U(n, f) can be reasonably well
reproduced by the calculated driving potential. We also learn that
the deformations of nuclei play an important role for a reliable
calculation of the driving potential. With a systematic study on
the reactions $^{238}$U(n, f), $^{237}$Np(n, f), $^{235}$U(n, f),
$^{232}$Th(n, f) and $^{239}$Pu(n, f), we find that the
experimental data of these reactions can be reproduced reasonably
well with the proposed fission potential. This investigation is
helpful for further describing the yields of the fission products.
By combining the radial basis function approach \cite{Wang11}, the
accuracy and predictive power of the model could be significantly
improved. In addition, a more microscopic description of the
potential parameters and the temperature dependence of the driving
potential should be further investigated. The study on these
aspects is under way.

\begin{center}
{\bf  ACKNOWLEDGMENTS}
\end{center}

We thank Min Liu and an anonymous referee for valuable
suggestions. This work was supported by Th-based Molten Salt Reactor Power System of Strategetic Pioneer Sci.\&Tech.
Projects from Chinese Academy of Sciences, Defense Industrial Technology Development
Program (No. B0120110034) and National Natural Science Foundation of China (Nos 10875031, 10847004 and 11005022).


\end{document}